\documentclass[twocolumn,preprintnumbers,superscriptaddress,nofootinbinb,aps,prd,floatfix]{revtex4-1}

\usepackage{amsmath}
\usepackage{multirow}
\usepackage[utf8]{inputenc}
\usepackage{graphicx}
\usepackage{color}
\begin{document}

\preprint{HU-EP-19/21}
\title{Exploring BSM Higgs couplings in single top-quark production}
\author{\textsc{Manfred Kraus, Till Martini, Sascha Peitzsch, Peter Uwer}}
\affiliation{\vspace{0.3cm}
 Humboldt-Universit\"at zu Berlin, Institut f\"ur Physik, Newtonstra\ss{}e 15, 
 12489 Berlin, Germany
}

\begin{abstract}
  In this article we study a Standard Model extension modifying the
  top-quark Yukawa coupling to the Higgs boson by allowing a mixture
  of CP-odd and -even couplings.  Single top-quark production in
  association with an additional Higgs boson provides a natural
  laboratory to search for such extensions. However,
  because of the small cross section the experimental analysis is
  challenging. Already the measurement of the cross section for this
  process is highly non-trivial.  Furthermore, using only cross
  section measurements, a certain parameter region would escape
  detection. Using an explicit BSM scenario we show
  that employing the Matrix Element Method a
  precise measurement becomes feasible. Ignoring signal detection
  efficiencies an integrated luminosity of about $20$ fb$^{-1}$ would
  allow a discovery. Assuming signal detection efficiencies at the level of a
  few percent a potential signal could be established in the high
  luminosity phase of the LHC.
\end{abstract}

\maketitle

\section{Introduction}
The discovery of the Higgs boson by the ATLAS and CMS collaborations
\cite{Aad:2012tfa,Chatrchyan:2012xdj} in 2012 completed the Standard
Model.  In many subsequent measurements the Higgs boson properties
were studied, analysing a variety of different Higgs boson
production and decay processes. Even though no deviations from the
Standard Model predictions have been reported by the experiments so
far, there is a great interest in further studies of the Higgs Yukawa
sector as this is a crucial part of the Higgs mechanism---the Standard
Model mechanism to spontaneously break the electroweak symmetry. In
particular, since in many beyond the Standard Model (BSM) scenarios the
Higgs sector is modified.  Prominent examples are for instance, the
two-Higgs-doublet model (2HDM)~\cite{Branco:2011iw} and composite Higgs
models~\cite{Agashe:2004rs} which both predict modified Higgs
couplings and introduce CP-violating Yukawa interactions.

In the Higgs sector of the Standard Model the top quark plays a
special role because of its large Yukawa coupling
$(y_t = \sqrt{2}~m_t/v \sim 1)$. Thus, if physics beyond the Standard
Model modifies the Higgs sector, deviations are expected to be seen
first in the top-Higgs interaction. This makes precise studies of the
top-Higgs dynamics a promising laboratory to search for new physics.

In fact, not only the value of the top-quark Yukawa coupling is of
high interest but also its relative sign with respect to the Higgs
boson interactions with gauge bosons. A relative sign between these
couplings different from the Standard Model prediction, has a severe
impact on the electroweak symmetry breaking mechanism and leads to a
violation of unitarity~\cite{Appelquist:1987cf}. As a further test of
the Standard Model, it is thus important to analyse the top-Higgs
coupling in more detail. Such studies can also provide a further
consistency test of the CP properties of the Higgs boson.

At the LHC, the top-quark Yukawa coupling is experimentally accessible only in 
a few processes. It can be inferred indirectly in the dominant Higgs 
boson production process via gluon fusion, where the Higgs couples to a closed 
top-quark loop. Recently, also an indirect measurement of the top-quark Yukawa 
coupling from electroweak corrections to top-quark pair 
production~\cite{Sirunyan:2019nlw} has been presented. 
Common to all indirect measurements is that
the extraction of the Yukawa coupling can be spoiled by unaccounted BSM effects.
In contrast, the production of a Higgs boson in association with 
top quarks allows a direct measurement of the corresponding interaction.
The recent discovery of the $t\bar{t}H$ process at the LHC
\cite{Sirunyan:2018hoz,Aaboud:2018urx} allows for the first time a direct 
determination of the interaction strength. However, neglecting the
masses of the light quarks,  the $t\bar{t}H$ cross section
is insensitive to the sign of the top-quark Yukawa coupling because of its
quadratic dependence. 

Complementary information can be gathered from Higgs boson production
in association with a single top quark. In particular, this process
offers the rather unique opportunity to study the relative sign
between the top-quark Yukawa and the Higgs-gauge boson interaction
terms.
\begin{figure}[h!]
\begin{center}
  \includegraphics[width=0.22\textwidth]{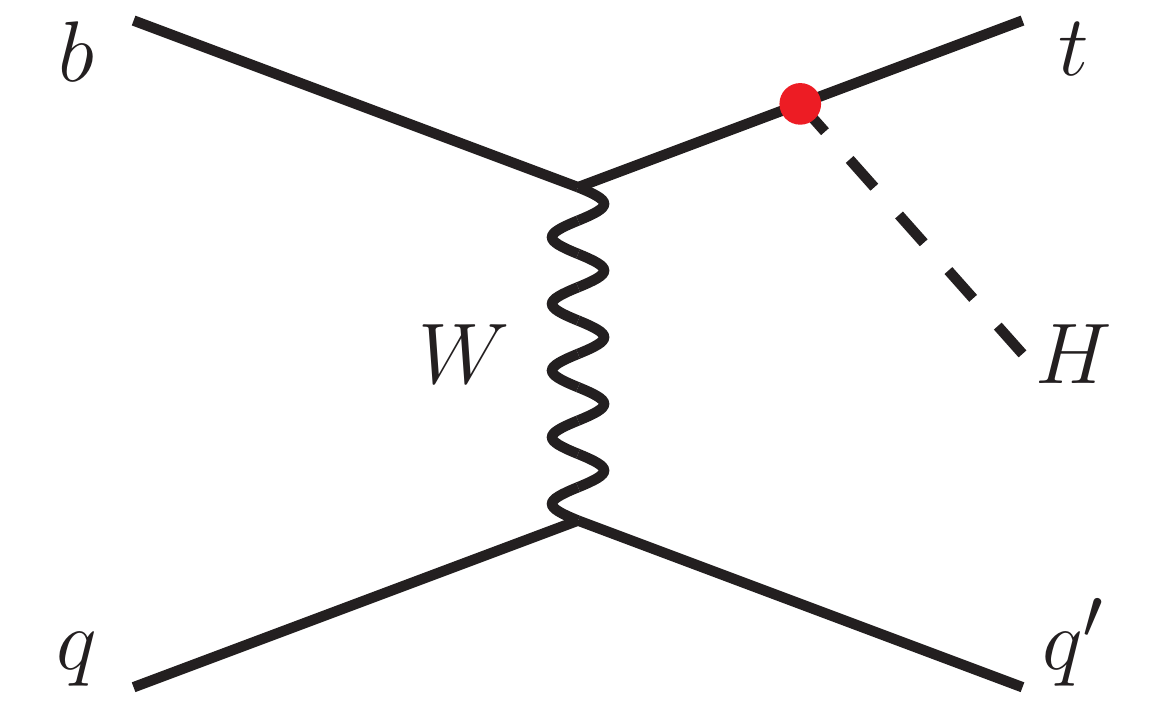}
  \includegraphics[width=0.22\textwidth]{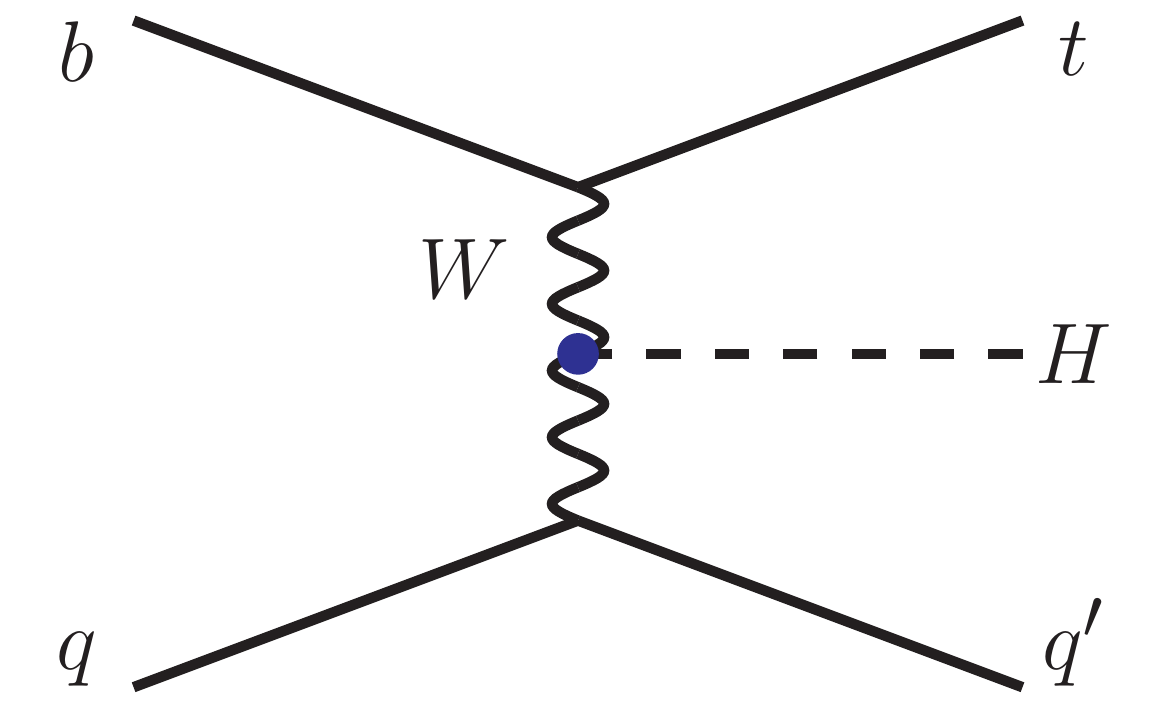}
\end{center}
\caption{\it  
 Leading order Feynman diagrams for $pp \to tH$ in the $t$-channel production 
 process.
}
\label{fig:diags}
\end{figure}

Within the Standard Model (SM) the dominant production mode at the LHC is
via the $t$-channel exchange of a $W$ boson, as illustrated in
Fig.~\ref{fig:diags}.  The Higgs boson can be emitted either from the
top quark or from the virtual $W$ boson. In the Standard
Model each individual contribution is sizeable. However, the total
cross section is reduced by a significant destructive interference between
the two diagrams, which originates from the cancellation of the
longitudinal $W$ boson polarization
\cite{Bordes:1992jy,Maltoni:2001hu,Farina:2012xp}. Thus, besides the
prominent example of longitudinal $W$-boson scattering, single
top-quark production in association with a Higgs boson also serves as
a testing ground for the unitarisation of the Standard Model as
realised by the Higgs mechanism.

Based on these unique opportunities the process has received a lot of
attention in recent
years~\cite{Bordes:1992jy,Maltoni:2001hu,Farina:2012xp,
  Biswas:2012bd,Agrawal:2012ga,Taghavi:2013wy,Campbell:2013yla,Ellis:2013yxa,
  Chang:2014rfa,Kobakhidze:2014gqa,Wu:2014dba,Englert:2014pja,Demartin:2015uha,
  Gritsan:2016hjl,Rindani:2016scj,Barger:2018tqn,Degrande:2018fog}
even though the experiments have only presented searches for this
production channel~\cite{Khachatryan:2015ota,Sirunyan:2018lzm} so far. The
reason is that the experimental signature is very challenging and the
total cross section is very small, requiring sophisticated analysis
techniques to become sensitive to this reaction.

In this article, we extend previous studies and use the Matrix Element
Method \cite{Kondo:1988yd,Kondo:1991dw} to extract the top-quark
Yukawa coupling. In recent years, the Matrix Element Method has been
established as a powerful method to extract information in cases where
one suffers from small event rates and complicated
backgrounds.
As a concrete example we investigate the Standard Model
extension studied in Ref.~\cite{Artoisenet:2013puc}. This model introduces
a CP-violating top-quark Yukawa coupling keeping at the same time the
top-quark mass fixed. It is thus possible to parametrise potential
deviations from the Standard Model in a continuous way. Alternatively,
one could also study a two-Higgs-doublet model as a specific example.

We give first some details on the computation and illustrate
the impact of the new coupling at the inclusive level. In addition, a
brief review of the general idea of the Matrix Element Method is
presented. Finally, we discuss the potential constraining power of the
method in the context of the $pp \to tH$ production process and
conclude.

\section{CP-violating Yukawa coupling}
\label{sec:yt}
In order to study the top-quark Yukawa coupling independently of the
top-quark mass we follow Ref.~\cite{Artoisenet:2013puc} and allow for
a mixture of CP-even and CP-odd interactions. In this scenario the
parametrization of the interaction vertex of the top quark and the
Higgs boson is given by \cite{Artoisenet:2013puc}
\begin{equation}
\mathcal{L}_{t\bar{t}H} = - \frac{y_t}{\sqrt{2}}\left(a\cos(\alpha)\bar{t}t + 
  ib\sin(\alpha)\bar{t}\gamma_5t\right)H.
\label{eqn:LttH}
\end{equation}
In addition, we use for the parameters $a$ and $b$ the setting of
Ref.~\cite{Demartin:2015uha}:
\begin{equation}
a = 1, \qquad b = \frac{2}{3},
\end{equation}
to keep the $gg\to H$ total cross section unmodified for any value of
$\alpha$.  Thus, the model has only one free parameter, namely the
CP-mixing angle $\alpha$. Varying $\alpha$ allows to interpolate
continuously between the CP-even $(\alpha=0^\circ)$ and the CP-odd
$(\alpha=180^\circ)$ scenario.

In the following, we study the impact of the modified coupling on the
dominant $t$-channel production of a single top quark in association with a Higgs 
boson at the next-to-leading order (NLO) in QCD.
The analysis is performed for the LHC Run II energy of $\sqrt{s}=13$ TeV. 
The following Standard Model input parameters are used
\begin{equation*}
 m_t = 173.2~\textrm{GeV}, \quad m_W = 80.385~\textrm{GeV}, \quad 
 m_H = 125~\textrm{GeV}.
\end{equation*}
We require that the hardest light jet, as well as the top-tagged jet
and the Higgs boson, fulfill the following selection cuts
\begin{equation*}
 p_T(X) > 30~\textrm{GeV}\;, \quad |\eta(X)| < 3.5\;, \quad X \in \{t,H,j\}\;,
\end{equation*}
where jets are defined using the $k_T$-jet
algorithm~\cite{Cacciari:2011ma} with a jet radius parameter of
$R=0.4$. In addition, we require that all final state objects are well separated
with $\Delta R_{ij} > 0.4$.The renormalisation and factorisation scales are set to the
common value of $\mu_R=\mu_F=m_t$. We have cross-checked the calculation
at the differential level with results obtained from
aMC@NLO~\cite{Alwall:2014hca,Artoisenet:2013puc}.
\begin{figure}[ht!]
\begin{center}
  \includegraphics[width=0.5\textwidth]{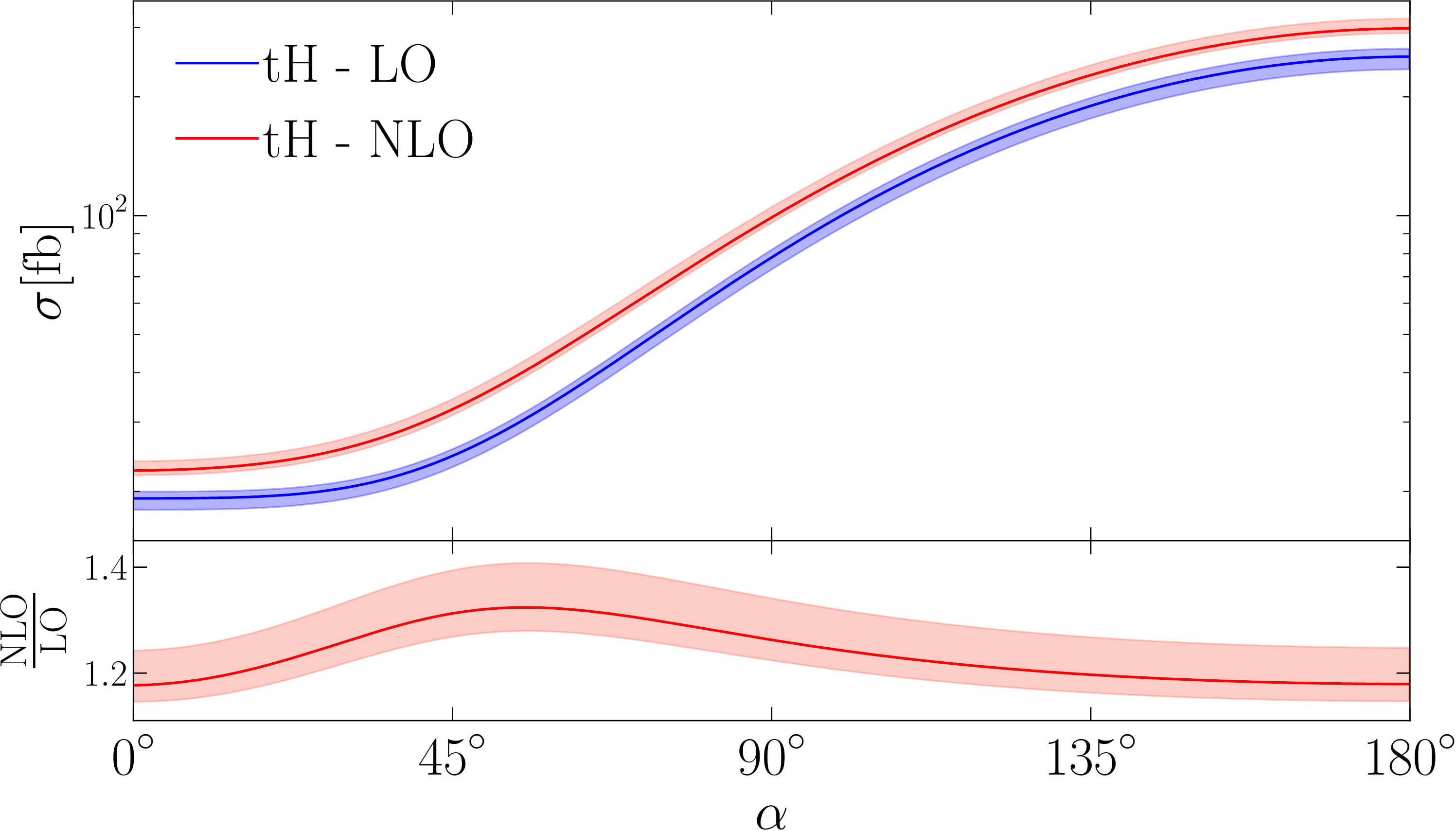}
\end{center}
\caption{\it  
 The fiducial cross section of $pp \to tH$ as a function of the CP-mixing 
 angle $\alpha$ at LO and NLO accuracy. 
}
\label{fig:xsec}
\end{figure}
In Fig.~\ref{fig:xsec} the fiducial cross section as a function of the 
CP-mixing angle $\alpha$ is shown. The upper panel depicts the predictions
for LO and NLO together with the theoretical uncertainties obtained by
varying the factorisation and renormalisation scale by a factor of two
up and down. The bottom panel illustrates the relative size of 
the NLO corrections with respect to the LO predictions.
The central prediction for the fiducial cross section at the next-to-leading 
order in QCD as a function of the CP-mixing angle $\alpha$ 
can be parametrised for the aforementioned setup by
\begin{multline}
 \sigma_{\textrm{fid}}^{\textrm{NLO}}(\alpha) = \sigma_{\textrm{fid}}^{\textrm{SM}}~\big[4.37 - 6.10\cos(\alpha) + 2.73\cos^2(\alpha) \\ 
	- 0.01\sin(\alpha) + 0.00\cos(\alpha) \sin(\alpha)\big]\;,
\end{multline}
with $\sigma_{\textrm{fid}}^{\textrm{SM}} = 22.6$ fb. The tiny value of the 
$\sin(\alpha)$ contribution and the compatibility of the $\cos(\alpha)\sin(\alpha)$ 
coefficient with zero (within its numerical uncertainty) are a consequence 
of the applied phase-space cuts, which constrain the preferred phase-space 
region of the pseudoscalar contribution~\cite{Demartin:2015uha}.
We want to stress that the smallness of the $\sin(\alpha)$ 
and $\cos(\alpha)\sin(\alpha)$ coefficients does not mean that the 
CP-odd contributions are negligible. In fact, contributions of the form 
$\sin^2(\alpha)$ are distributed across the constant and the 
$\cos^2(\alpha)$ coefficients.
For completeness we also provide the total inclusive cross section at 
the next-to-leading order in QCD
\begin{multline}
 \sigma_{\textrm{tot}}^{\textrm{NLO}}(\alpha) = \sigma_{\textrm{tot}}^{\textrm{SM}}~\big[4.70 - 5.19\cos(\alpha) + 1.49\cos^2(\alpha) \\ 
	- 0.95\sin(\alpha) + 0.59\cos(\alpha) \sin(\alpha)\big]\;,
\end{multline}
with $\sigma_{\textrm{tot}}^{\textrm{SM}} = 46.7$ fb.
Notice that due to the coupling structure given by Eqn.~\eqref{eqn:LttH} a 
similar decomposition holds for all observables for this process as long as 
only QCD corrections are taken into account.

Inspecting Fig.~\ref{fig:xsec}, one observes that the NLO corrections
are positive and sizeable over the whole range. The Standard Model
prediction as well as the `inverted' SM $(\alpha = 180^\circ)$ receive
corrections up to $+18\%$, while for $\alpha \approx 56^\circ$ the
corrections can reach up to $+32\%$. Note, that the maximum of the NLO 
corrections happens to coincide with
the maximal mixing of the CP-even and CP-odd coupling, which is
different from $\alpha = 45^\circ$ due to the choice $a \neq b$ in
Eqn.~\eqref{eqn:LttH}. As one would naively expect, the theoretical
uncertainty, as estimated by scale variation, is rather constant over
the whole range and amounts to $+6\%$ and $-3\%$. The fact that the
NLO prediction lies outside of the LO uncertainty band underlines the
importance of the NLO calculation.  The non-overlapping uncertainty
bands are not surprising since in leading order the process $pp\to
tH$ is of pure electroweak nature.  Thus, it is mandatory to
include NLO corrections since they are large and depend on
the CP-mixing angle $\alpha$.
\begin{figure}[h!]
\begin{center}
  \includegraphics[width=0.5\textwidth]{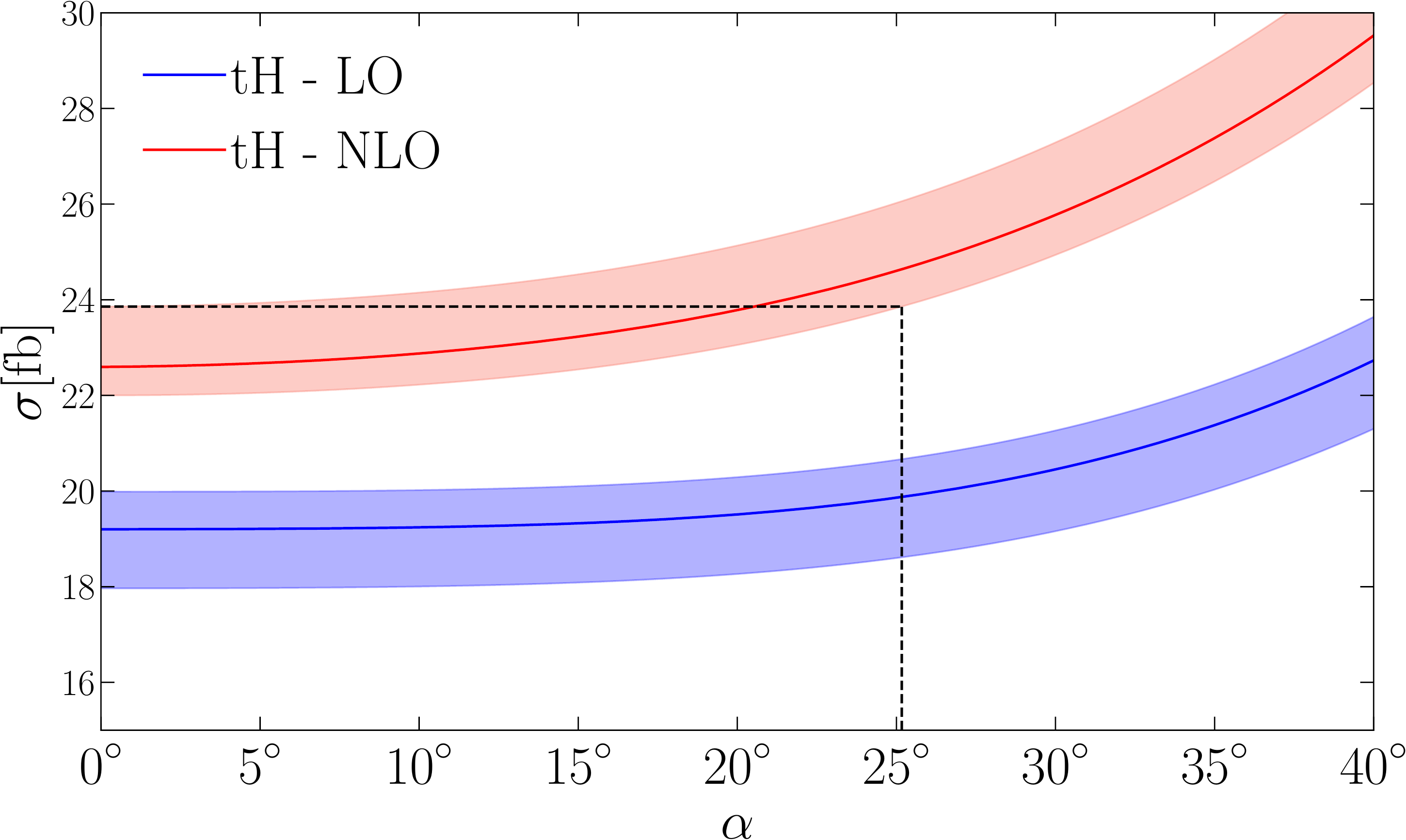}
\end{center}
\caption{\it  
 Zoom of Fig.~\ref{fig:xsec}. The dashed lines illustrate the interval
 in $\alpha$ that  is still compatible with the SM prediction within its 
 uncertainties.
}
\label{fig:xsec_zoom}
\end{figure}

Zooming into the region of small $\alpha$ (c.f. Fig.~\ref{fig:xsec_zoom}), one 
can see that using only the information of the inclusive cross section a 
significant interval of $\alpha$ cannot be distinguished from the Standard 
Model case ($\alpha=0^\circ$). In fact, by taking the scale uncertainty into 
account, the cross sections obtained for $0^\circ \le \alpha \le 25.2^\circ$ are 
compatible with the Standard Model prediction.

A more precise determination of the CP-mixing angle $\alpha$ is
expected from shape differences of differential distributions.
However, using standard cut-based techniques this requires a
sufficient amount of recorded events which will only become available
in the far future. Another drawback of differential distributions is
that detailed studies~\cite{Demartin:2015uha} are necessary in order to isolate 
specific observables that show a good sensitivity on the parameter under
scrutiny. If on the other hand several observables are used at the same
time, possible correlations, which can be difficult to quantify, need
to be taken into account.  In order to get the full benefit of
currently available data, we propose to use the Matrix Element Method. In
the next section, we give a brief introduction before applying the
method to the concrete case.

\section{The Matrix Element Method}
The Matrix Element Method (MEM) was first introduced in Refs.
\cite{Kondo:1988yd,Kondo:1991dw}  to separate signal from background
processes. However, the method is rather general and can also be used
to determine  
parameters of the underlying theory. While the formulation of the MEM has been 
restricted to LO accuracy for a long time, the inclusion of NLO QCD effects has 
been expedited within the last decade~\cite{Alwall:2010cq,Campbell:2012cz,Baumeister:2016maz} 
and has recently been achieved in a generic way~\cite{Martini:2015fsa,
Martini:2017ydu,Martini:2018imv,Kraus:2019qoq}. Given the large QCD
corrections observed in the previous section, incorporating NLO
corrections is mandatory to obtain reliable results.

In the case of the determination of a single parameter $\alpha$, the 
application of the MEM amounts to the following steps in practice. For each 
event of the data sample,
described by some set of kinematical variables which we collectively call 
$\mathbf{x} = \{x_1,...,x_n\}$, a weight is calculated under a specific 
assumption on the value of $\alpha$ according to
\begin{equation}
 P(\mathbf{x}|\alpha) = \frac{1}{\sigma(\alpha)}\int d^ny~ 
 \frac{d^n\sigma(\mathbf{y}|\alpha)}{dy_1...dy_n}~W(\mathbf{y},\mathbf{x}).
 \label{eqn:wgt}
\end{equation}
Here $W(\mathbf{y},\mathbf{x})$ is the so-called \textit{transfer
  function} that incorporates the unfolding of the detector signature
$\mathbf{x}$ back to the theoretically modelled final state
$\mathbf{y}$. The weights in Eqn.~\eqref{eqn:wgt} define a probability
density with
\begin{equation}
 \int d^nx~P(\mathbf{x}|\alpha) = 1,
\end{equation}
as long as the transfer function $W(\mathbf{y},\mathbf{x})$ is normalized to
one. This allows for a unique statistical interpretation within the method 
of Maximum Likelihood. Here, the negative log-Likelihood function is 
defined by
\begin{equation}
 -\log(\mathcal{L}) = -\sum_{i=1}^N \log(P(\mathbf{x}_i|\alpha)),
\end{equation}
where the sum runs over all $N$ events in the data sample. 
The estimator $\hat{\alpha}$ inferred from a given 
event sample is then obtained by minimising the negative log-Likelihood function
\cite{Cowan:1998}. In the following and as a first step, we will always assume 
a perfect detector and set 
\begin{equation}
  W(\mathbf{y},\mathbf{x}) = \delta(\mathbf{y}-\mathbf{x}).
\end{equation}
In the present case we use the energies, pseudo rapidities and azimuthal angles 
of the hardest light jet and the Higgs boson and the pseudo rapidity of the 
top-tagged jet as variables to describe the
events: 
\begin{equation}
 \mathbf{x}=\{\eta_t,E_j,\eta_j,\phi_j,E_H,\eta_H,\phi_H\}\;.
\end{equation}
In case of angular variables, modelling the transfer functions via delta
functions is considered a good approximation. For variables
related to energies of jets a more realistic description should include the
modelling of non-trivial jet-energy scales.

Even though the MEM can become computationally demanding for
non-trivial transfer functions $W(\mathbf{y},\mathbf{x})$ the method
has already proven its strength in several applications where only
limited amount of data was available.  Prime examples of its success
are the determination of the top-quark mass
\cite{Abbott:1998dn,Abazov:2004cs,Abulencia:2006ry} and the discovery
of the single top-quark
production~\cite{Abazov:2009ii,Aaltonen:2009jj} at the
Tevatron. Furthermore, at the LHC evidence for the $s$-channel
production of single top quarks has been found by virtue of the
MEM~\cite{Aad:2015upn}.  It is in this domain, where the multi-variate
MEM can overcome the abilities of traditional methods.
\section{Results}
Based on the formalism introduced in Ref.~\cite{Martini:2015fsa,
Martini:2017ydu,Martini:2018imv,Kraus:2019qoq} we consider the
aforementioned BSM scenario with 
a non-vanishing CP-mixing angle. To analyse the potential of the
method, we simulate a measurement by  generating unweighted NLO events for a 
CP-mixing angle of $\alpha = 22.5^\circ$ and a renormalisation and factorisation
scale of $\mu = m_t$.  These events serve as pseudo-data in 
the following analysis. The remaining setup remains the same as in 
Section~\ref{sec:yt}. As discussed in Section~\ref{sec:yt} such a deviation
from the SM cannot be resolved relying only on the measurement of the
inclusive cross section. In the following, we analyse to which
extent the MEM is able to recover the input value for the CP-mixing
angle and establish a deviation from the SM.

We always use Likelihood functions based on fixed-order NLO weights,
since it has been shown in previous
studies~\cite{Martini:2015fsa,Martini:2017ydu, Martini:2018imv} that
the NLO description significantly improves the parameter determination
and the reliability of the uncertainty estimates; even if parton
shower effects are included in the pseudo-data~\cite{Kraus:2019qoq}.

\begin{figure}[h!]
\begin{center}
\includegraphics[width=0.5\textwidth]{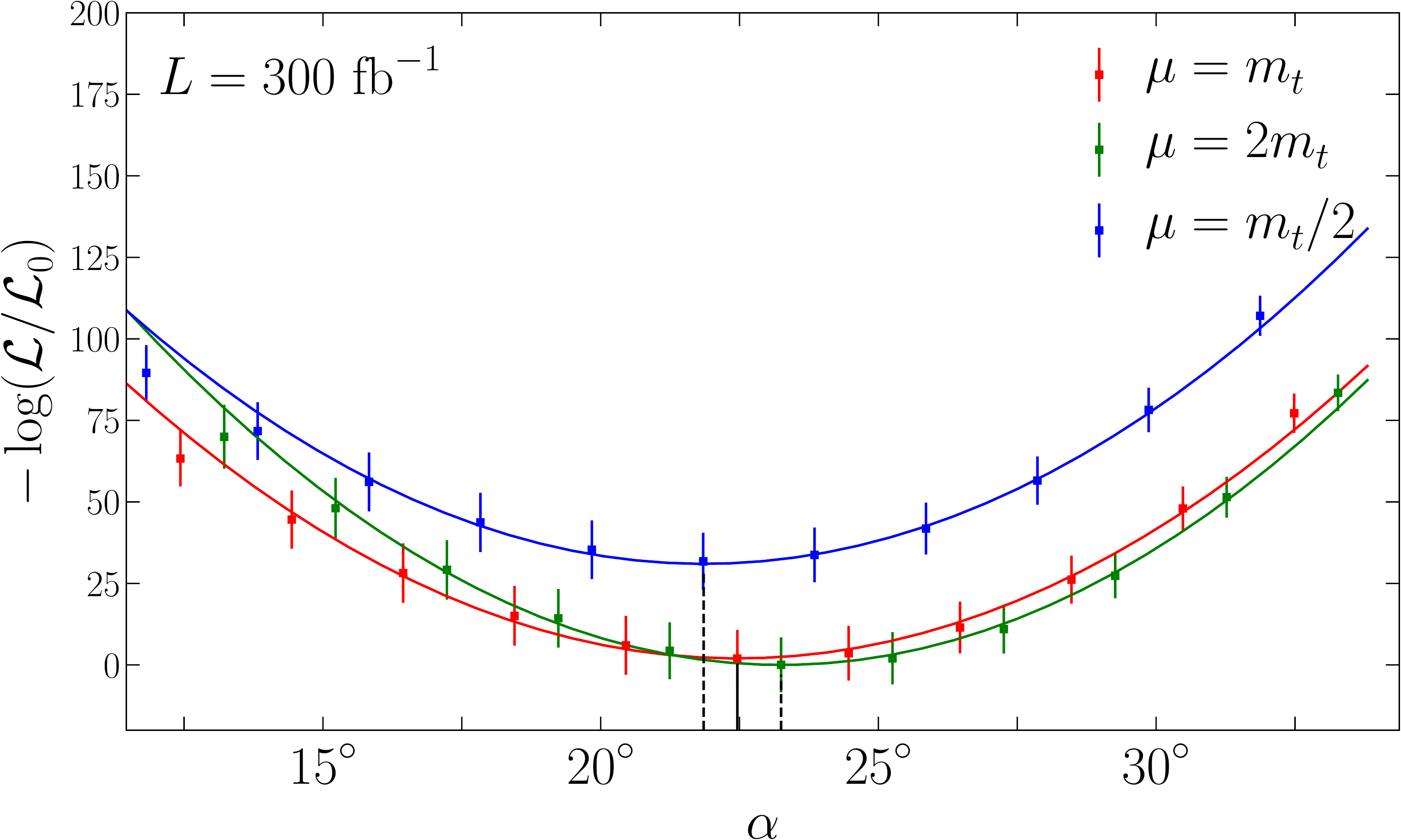}
\end{center}
\caption{\it The minima of the negative log-Likelihood functions for three 
 different scale choices: $\mu = m_t/2$, $\mu = m_t$ and $\mu = 2m_t$.
 The pseudo-experiment assumes an integrated luminosity of $300$ fb$^{-1}$.}
\label{fig:fit}
\end{figure}

\begin{figure}[h!]
\begin{center}
\includegraphics[width=0.5\textwidth]{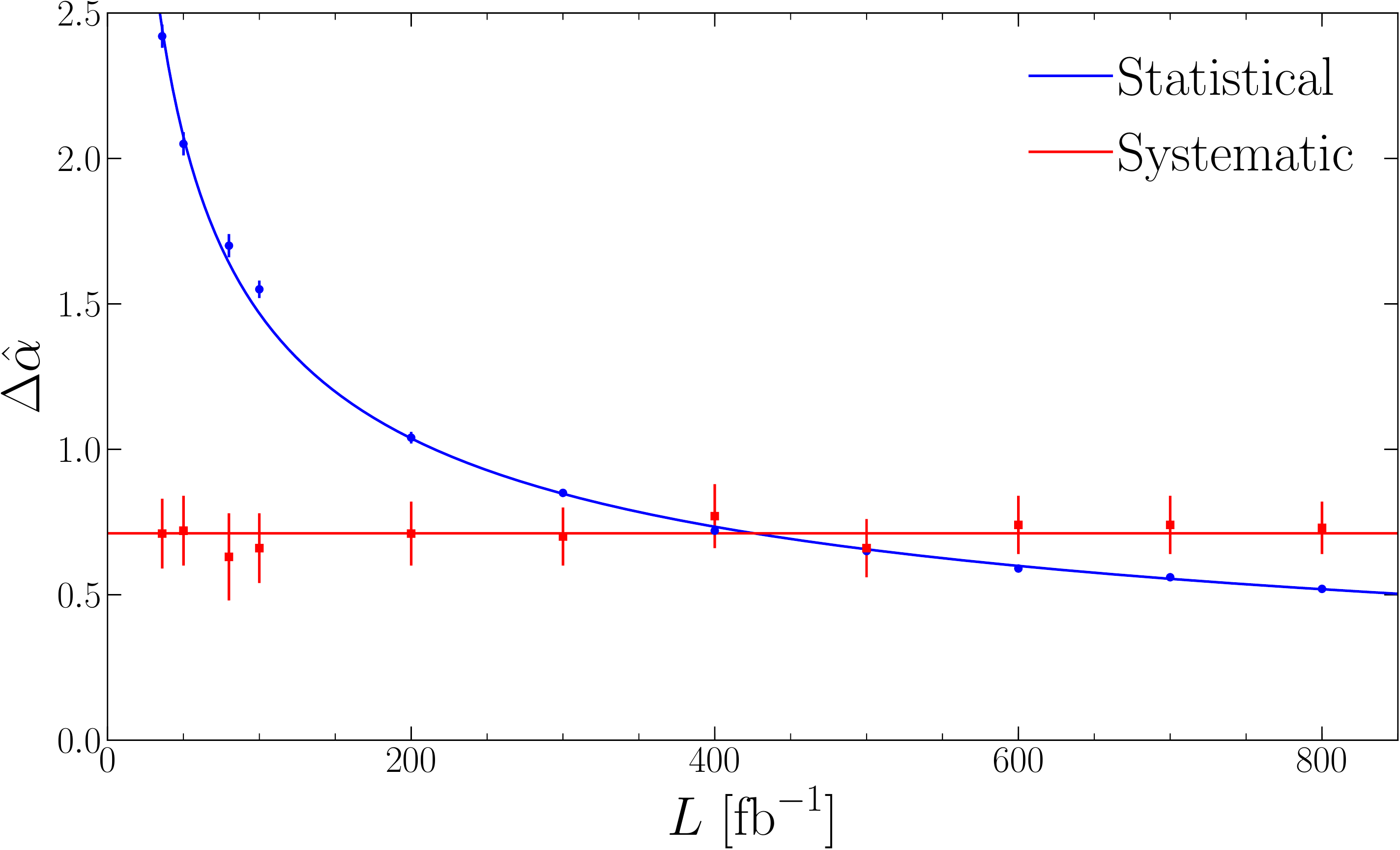}
\end{center}
\caption{\it The estimated statistical and systematic uncertainty, 
 $\Delta\hat{\alpha}$ on the extracted CP-mixing angle $\hat{\alpha}$ as a 
 function of the integrated luminosity $L$. The solid lines show a constant and
 a $1/\sqrt{L}$ fit.}
\label{fig:uncert}
\end{figure}

In Fig.~\ref{fig:fit} the negative log-Likelihood function is shown
for three different scale choices and an assumed integrated luminosity
of $300$ fb$^{-1}$, while no signal detection efficiencies have been taken into account.
The central scale, $\mu = m_t$, is shown in red, while $\mu = 2m_t$ is depicted
in green and $\mu = m_t/2$ in blue. The estimator for the CP-mixing angle 
$\hat{\alpha}$ is determined from the position of the minimum of the 
Likelihood function for $\mu = m_t$, while the statistical uncertainty 
$\Delta\hat{\alpha}_{\textrm{stat}}$ is deduced from the width of the 
fitted parabola. The Likelihood functions based on the other two scales allow
to estimate the systematic uncertainty $\Delta\hat{\alpha}_{\textrm{sys}}$ by
studying the impact of higher-order corrections on the extracted angle.
For the assumed $300$ fb$^{-1}$ we obtain as estimator for the CP-mixing angle:
\begin{equation}
 \hat{\alpha} = 22.5^\circ \pm 0.9^\circ~\textrm{[stat.]}~^{+0.8^\circ}
 _{-0.6^\circ}~\textrm{[sys.]}\;,
\end{equation}
where the statistical uncertainty is still the dominant error contribution.
In Table~\ref{tab:fit} we also show results for $L=36$ fb$^{-1}$ and 
$L=80$ fb$^{-1}$, together with the previously discussed result for $L=300$ 
fb$^{-1}$.
\begin{table}[h!]
\begin{center}
\begin{tabular}{ccccc}
 \hline
 $L~[\textrm{fb}^{-1}]$ & $\hat{\alpha}$ & $\Delta\hat{\alpha}_{\textrm{stat}}$ & 
 $\Delta\hat{\alpha}^+_{\textrm{sys}}$ & $\Delta\hat{\alpha}^-_{\textrm{sys}}$ \\
 \hline
 $36$  & $23.2^\circ$ & $\pm 2.4^\circ$  & $+0.5^\circ$ & $-0.7^\circ$ \\[0.1cm]
 $80$  & $24.2^\circ$ & $\pm 1.7^\circ$  & $+0.3^\circ$ & $-0.9^\circ$ \\[0.1cm]
 $300$ & $22.5^\circ$ & $\pm 0.9^\circ$  & $+0.8^\circ$ & $-0.6^\circ$ \\[0.1cm]
 \hline
\end{tabular}
\caption{\it The inferred value of the CP-mixing angle and its statistical and 
 systematic uncertainties for several integrated luminosities.}
\label{tab:fit}
\end{center}
\end{table}
As expected, the statistical uncertainty reduces with increasing luminosity, 
while the systematic uncertainty is independent of the number of events.
We can test this expectation by studying the uncertainty contributions as a 
function of the luminosity, as illustrated in Fig.~\ref{fig:uncert}.
The red square data points represent the systematic uncertainties
inferred from the Likelihood analysis while the blue dots correspond
to the statistical uncertainty.  Note that for this study we used the
symmetrised version of the systematic uncertainties. The solid lines
depict the fit results assuming a constant for the systematic uncertainties
and a $1/\sqrt{L}$ behaviour for the statistical uncertainty. We
conclude that the systematic uncertainty inferred from scale
variations amounts to a constant $\pm 0.7^\circ$ uncertainty.
Starting from an integrated luminosity of $L\approx 425$ fb$^{-1}$
onwards, the analysis would not improve any more by collecting more data
and is limited by the theory uncertainty. However, assuming realistic signal detection
efficiencies at the level of a few percent~\cite{Sirunyan:2018lzm}, even for the high
luminosity phase of the LHC the NLO predictions would be sufficiently precise.
In the long term perspective a determination of the mixing angle could
envisaged with an absolute accuracy of $\pm 0.7^\circ$.

To conclude, we want to give an estimate for the required integrated luminosity 
in order to achieve a discovery, i.e. a $5\sigma$ deviation from the SM value
$(\alpha = 0^\circ)$, at the LHC for the BSM scenario under 
consideration. For simplicity we assume that the SM value can be inferred
with the same accuracy as the BSM scenario studied before. 
We find that approximately $20$ fb$^{-1}$ are necessary to establish a $5\sigma$
deviation from the Standard Model. Assuming a realistic signal detection efficiency of 
$3\%$ an integrated luminosity of $300$ fb$^{-1}$ allows to establish an excess with 
$3\sigma$ significance. However, even with this realistic signal detection efficiency a
discovery would be possible in the high-luminosity phase of the LHC.
\section{Conclusions}
In this article we study a Standard Model extension which generalizes
the top-quark Yukawa coupling to the Higgs boson.  In particular, the
model allows that the Higgs boson is not a CP eigenstate and
interpolates continuously between the SM and a scenario where the
relative sign of the top-quark Yukawa coupling with respect to the
Higgs boson coupling to $W$ bosons is inverted. We find that for mixing
angles up to $25^\circ$ a cross section measurement alone cannot
distinguish the SM from the BSM scenario. 

Simulating a concrete measurement and assuming a signal detection efficiency 
of $3\%$ we show that using the Matrix Element Method a signal at the $3\sigma$ level can be established with 
an integrated luminosity of $300$ fb$^{-1}$.
The high luminosity phase allows a discovery with $5\sigma$ and will not be 
limited by theoretical uncertainties. As a side effect the paper provides a 
further illustration of the power of the Matrix Element Method in challenging 
the Standard Model.

\section*{Acknowledgments}
This work is supported by the German Federal Ministry
for Education and Research (grant 05H15KHCAA).

\end{document}